\documentclass[12pt]{iopart}
\usepackage{graphicx}

\begin{document}

\title{Halos and related structures}

\author{K Riisager}

\address{Department of Physics and Astronomy,
Aarhus University, Ny Munkegade 120, DK-8000 Aarhus C}
\ead{kvr@phys.au.dk}

\begin{abstract}
The halo structure originated in nuclear physics but is now
encountered more widely. It appears in loosely bound, clustered
systems where the spatial extension of the system is significantly
larger than that of the binding potentials. A review is given on our
current understanding of these structures, with an emphasis on how the
structures evolve as more cluster components are added, and on the
experimental situation concerning halo states in light nuclei.
\end{abstract}

\submitto{\PS}
\maketitle

\section{Introduction}
The word ``halo'' is used in many areas of human activity as a glance
in an encyclopaedia will show. A common feature to most of these
meanings is an extended peripheral distribution, most often dilute,
around a central object. This also holds for the object of this
contribution: the quantum halos found up to now mainly in nuclear
physics, but also known to exist in molecular physics. A more precise
definition follows in the next section, for now a spatial extension
significantly larger than that of otherwise similar systems will do as the
characteristic property. They first came into focus through the
measurements of nuclear matter radii carried out by Tanihata and
collaborators \cite{Tan85a,Tan85} and obtained their name 25 years ago
in the paper by Hansen and Jonson where the key ingredient of their
structure was identified \cite{Han87}. Halos were a major contributor
to the growing interest in radioactive beams that followed shortly
after and aspects of halos and their study have been covered in many
review papers since then, among them
\cite{Ber93,Zhu93,Rii94,Tan95,Han95,Tan96,Jon98,Jen01,Jon04,Jen04,Rii06}.

It is only natural that halos will appear in many of the contributions
to these proceedings. The aim here is to give an overview of what
characterizes the halo systems and how they are probed
experimentally. Section \ref{sec:struc} focuses on the halo structure,
what distinguishes it from other systems as well as its intimate
relation to the Efimov effect. Section \ref{sec:probe} gives a quick
overview of the many types of experiments that have been used to study
nuclear halos and also attempts to extract generic traits. The final
section \ref{sec:out} summarizes which pronounced halos are
established currently and points to some of the yet unsolved
questions.

\section{The structure of halos}  \label{sec:struc}
\subsection{Basics}
The understanding of halos that has emerged after the first two
decades of study has changed very little since the latest major review
\cite{Jen04} and I shall rely on that for much of the general
description and refer to it for a more detailed exposition.

The defining feature of a halo was from the beginning understood to be
a large spatial extension caused by neutrons tunneling out from a
nuclear core. This picture, based on the first established cases such
as $^{11}$Li, still prevails but must be refined in order to use the
concept consistently also in other physics disciplines
\cite{Jen04,Rii00}.  It is not sufficient that a system is large, the
tunneling that arises from the quantum wave nature must also be a
prominent feature of the system in order that we obtain structures
that are truly universal. As an example Rydberg atoms are therefore
excluded since most of the wavefunction here resides in a classically
allowed region, a long-range attractive potential will in general not
give halos. The system should be divideable into a core and one or
more halo particles that can tunnel out making cluster models or
few-body models a natural first choice for describing halos. Tunneling
will occur in all quantum systems but it should be significant before
one can expect a system to reach the regime of universal
structure. One counter-example is the periphery of heavy nuclei where
the neutron density will extend further out than the proton density
--- seen e.g.\ elegantly through antiprotonic $^{208}$Pb and
$^{209}$Bi atoms \cite{Klo07} (the term ``halo'' is unfortunately also
employed traditionally in those studies, but the physics is of course
quite different) --- without having any significant dynamical effect
on the overall system in contrast to light halo nuclei.

A few more technical comments may be useful before proceeding, a more
complete discussion can again be found in \cite{Jen04}.  A short-range
potential is one that falls off with distance more rapidly than
$r^{-2}$, i.e.\ $r^2V(r) \rightarrow 0$ for $r \rightarrow
\infty$. The obvious measure of the range of the potential, given a
halo particle of a certain binding energy, is the classical turning
point $R$ of the particle where its potential energy equals its total
energy. This definition assumes we are dealing with a two-body system,
the case of a two-body subsystem of a many-body system is not so
straightforward. In practice potentials fall off sufficiently rapidly
that we may approximate $R$ by using quantities that are more easily
accessible experimentally. For two-body halo nuclei one can use the
equivalent square-well radius that is related to the mean-square radii
of the two components by $3/5 R^2 = \langle r^2 \rangle_1 + \langle
r^2 \rangle_2 + 3.3$ fm$^2$, where the last term reflects the finite
range of the nuclear force. The good halos found early on in nuclear
physics, such as $^{11}$Li and $^{11}$Be, have probabilities of being
clustered and of having halo particles outside of $R$ that are all
above 50\% and this value could be used as a yardstick for singling
out well-developed halo systems.

Another question that will return in the next subsection is how one in
practice decides how many ``clusters'' a halo state should be divided
into. For most systems this is not a practical problem, but a
challenging example is given by the hypertriton, $^3_{\Lambda}$H =
n+p+$\Lambda$, that is bound by about 0.14 MeV with respect to
break-up into d+$\Lambda$ but where the deuteron in turn is bound by
only 2.2 MeV and therefore in itself is a quite good halo. Even though
three-body models must be used to give a good description of
$^3_{\Lambda}$H we shall see that it is most naturally considered a
two-body halo.

Now, if a system is well-clustered and can be described in terms of a
short-range potential a halo state will appear once the binding energy
is sufficiently small. The tricky part of the question of when halos
will form is therefore whether sufficient clustering is present. In
general a component can only be expected to remain inert if the energy
required to excite it or break it apart is higher than the interaction
energy between components. On the molecular scale, where atoms are the
natural ``building blocks'' this is almost automatically fulfilled and
one can expect molecular physics to provide interesting halo test
cases. On the atomic scale, where electrons are added as active
ingredients, the electrons in atom components are easily perturbed,
but one may look for halos in systems of the type alkali-atom + e +
nobel-gas-atom or electrons bound to a molecule with permanent dipole
moment. Atomic halos should be rare and no good case has been
identified in experiment so far. Finally, on the nuclear scale, where
nucleons could serve as halo particles around a core, the conditions
for clustering are less clear and therefore in a sense more
interesting. Nuclei may provide us not only with well-developed halo
states but also with systems intermediate between normal nuclei and
halo nuclei. The key question is to what extent the core will remain
inert. This subject is closely related to nuclear cluster physics
\cite{Oer06,Fre07} and we shall return to it in section \ref{sec:out}.
Further adding to their interest is that, as will also become clear in
section \ref{sec:probe}, halo states in nuclei are characterized by
their special dynamical behaviour.

There is as yet no clear answer to where nuclear halo states will
occur. Nevertheless several calculations have been made, the most
recent ones \cite{Rot09,Rot09a,Sch08} within Hatree-Fock-Bogoliubov
theory. The criteria for defining halos in heavier nuclei are
typically different from the ones given above, see the more extended
discussion in \cite{Rot09a}, but at least for even-even nuclei it
appears that halos will be less numerous (and on a relative scale less
extended) than for light nuclei. Part of this is due to pairing that
has an important, but intricate, effect on halos \cite{Rot09}, in
certain cases decreasing sizes by mixing orbitals, in others enhancing
them.  Some of the necessary general conditions have been identified
and will be mentioned in the next subsection. An important restriction
is the mixing with others states that cannot be avoided if the local
density of states is too high. This limits the occurence of halos in
excited states, the estimates in \cite{Jen00} for an excitation energy
$E^*$ use a level distance of $D_0\exp(-2\sqrt{aE^*})$ (with $D_0 = 7$
MeV and $a = A/7.5$ MeV) and conclude that s-wave neutron halo states
should have binding energy $B< 270\,\mathrm{keV}\,\left( A/Z
\right)^2\exp(-4\sqrt{aE^*})$ and that the restriction for p-waves
is $Z < 0.44 A^{4/3}\exp(-2\sqrt{aE^*})$. For now, the main inference
from these estimates is that nuclear halos will predominantly occur in
ground states or at low excitation energy and therefore is a dripline
phenomenon.

\subsection{N-body systems}
Several contributions to this symposium deal with the challenges posed
by describing nuclear systems. For halo systems the requirements are
to describe correlations in the systems well and to describe the large
distance behaviour accurately. Not all theoretical frameworks can do
this easily (as an example some work better in momentum space whereas
halos are described more naturally in configuration space ). I shall
focus here on lighter nuclei where few-body models can be used and
start by commenting on the classification of few-body systems.

Knot theory in mathematics gives a well-estalished classification of
linked systems including the famous Borromean rings: three linked
rings where each pair is unlinked. A general system of $N$ rings is
called Borromean if each subsystem of 2 rings is unlinked, and
Brunnian if all subsystems with $N-1$ rings is unlinked (this can be
generalized further \cite{Bas12}). Following \cite{Zhu92} this
nomenclature was taken over in physics by replacing ``geometrically
(un)bound in three dimension'' with ``energetically
(un)bound''. Although mainly used for three-body systems, one could
therefore speak of $N$-body Borromean systems if all two-body
subsystems are unbound. Examples of nuclear Borromean systems are
$^6$He ($\alpha$+n+n), $^9$Be ($\alpha$+$\alpha$+n), $^{11}$Li
($^9$Li+n+n) and $^{45}$Fe
($^{43}$Cr+p+p), but one can also find five-body Borromean nuclei such
as $^8$He and $^{19}$B that consist of a core nucleus and four
neutrons. Similar structures will be abundant for heavier dripline
nuclei.

One can gain a first overview of the possible behaviours of an
$N$-body system by going from the $3N$ spatial coordinates $\vec{r}_i$
to generalized hyperspherical coordinates \cite{Zhu93,Nie01}. After
separating out the three centre-of-mass coordinates $\vec{r_{cm}}$ a single
radial coordinate $\rho$ is defined through 
\begin{equation}  \label{eq:rho}
  m\rho^2 = \sum_i m_i (\vec{r}_i-\vec{r_{cm}})^2 =
  \sum_{i<k} \frac{m_im_k}{M} (\vec{r}_i-\vec{r}_k)^2
\end{equation}
where $ M = \sum_i m_i$ and $m$ is chosen as a typical mass scale of
the system (for nuclear halos: a nucleon mass). The remaining $3N-4$
coordinates are dimensionless and are generalized angles that
incorporates the information on relative distances in the system. From
the kinetic energy term one can extract a term only depending on
$\rho$ that appears in the same functional shape as a centrifugal
barrier:
\begin{equation}
  \frac{\hbar^2}{2m}\frac{\ell^*(\ell^*+1)}{\rho^2} \;,\; \ell^* =
  \frac{3}{2}(N-2) \;.
\end{equation}
Note that this contribution also appears for a system where all
particles are in relative s-waves, the effective angular momentum
given in table \ref{tab:effcen} therefore suggests that systems with
more particles will be more confined. This can be understood from the
following argument: if a wavefunction is allowed to extend away from
origo it will spread out much faster in a higher-dimensional space
(the volume increases more rapidly with distance) and the overlap with
the binding potential close to the origo will therefore decrease, to
counteract this and keep the total system bound it needs to stay
small. The obvious loophole in this argument is that correlations in a
subsystem may ensure binding, mathematically this corresponds to having
$\rho$ large but one (or more) of the distances $\vec{r}_i-\vec{r}_k$
small (giving important contributions also from the ``angle''-part of
the Hamiltonian). I shall look at this in detail below for
three-particle systems where one often introduces the hypermomentum
$K$ as a generalization of the two-body angular momenta, $K=0$
corresponding to the most symmetric state with s-waves in all
subsystems.

\begin{table}
  \caption{\label{tab:effcen}Effective angular momentum for an
    $N$-body system.}
  \begin{indented}
  \item[]\begin{tabular}{@{}lcccc}   \br
      $N$ & 2 & 3 & 4 & 5 \\   \mr
      $\ell^* $ & 0 & 3/2 & 3 & 9/2 \\ \br
    \end{tabular}
  \end{indented}
\end{table}

There is no unique best way to measure the spatial extent of a
halo. The tradition is to use the mean-square (or root-mean-square)
radius, this is with the above choice of coordinates given as
\begin{equation}
 M \langle r^2 \rangle = m \langle \rho^2 \rangle + \sum_i m_i \langle
 r^2 \rangle_i \;,
\end{equation}
where the last term contains the contributions from the size $\langle
r^2 \rangle_i$ of each of the components in the total system. This
measure of extent emphasizes the tails and will therefore be quite
sensitive to the dilute tails in halos. In several cases a more
appropriate measure is the probability of finding a particle outside a
certain distance (e.g.\ in a classically forbidden region) as we shall
see when discussing the experimental probes of halos.

Since an $N$-body system without internal correlations (all particles
in relative s-waves) appears as a two-body system with an effective
angular momentum, we can start by considering two-body systems in the
limit of small binding energy $B$. This can be solved in the general
case \cite{Rii92} and the expectation value of $r^n$ will go as
\begin{equation}  \label{eq:asymp}
  \langle r^n \rangle \sim \begin{array}{l}
    (\mu B)^{(2\ell-1-n)/2}  \\  \ln(\mu B)  \\ \mathrm{constant}  \end{array}
    \mathrm{for} \;\;n  
    \begin{array}{c} > \\ = \\ <  \end{array}  2\ell-1 \;,
\end{equation}
i.e.\ it diverges in the zero-energy limit unless $n<2\ell-1$. An
angular momentum of course confines a system, but this can be
compensated by a higher weighting of the tail region. From the above
analysis we therefore deduce that the mean square radius will go to
infinity for vanishing binding for two-body systems with relative s-
and p-waves and for three-body systems with $K=0$ whereas all other
systems remain finite. However, it should be noted that the only
really pathological system is the two-body s-wave where the
probability of being outside the binding potential can go to 1. This
does not happen in any other case.

If a long-range repulsive interaction is present, e.g.\ the Coulomb
field for nuclear proton halos, the tail of the wavefunction will be more
suppressed at larger distances than for an angular momentum
barrier. This makes halo formation more difficult and will prevent it
altogether for nuclear charges above 10--20 (none of the identified
one- or two-proton emitters will have halo character). On the other
hand it has been shown explicitly that nuclear deformation will not
prevent formation of halos \cite{Mis97}.

We are interested in being able to compare also systems at finite
binding. To do this across physics fields it is useful to introduce
dimensionless scaling variables \cite{Rii00,Jen03}. For a two-body
system we use the classical turning point that in practice is
approximated by the $R$ discussed above. The dimensionless mean square
radius is then simply $\langle r^2 \rangle / R^2$. The momentum
corresponding to a range $R$ is of order $\hbar/R$ giving a kinetic
energy of order $\hbar^2/(2\mu R^2)$, so a natural dimensionless
binding energy is therefore $\mu BR^2/\hbar^2$. With this set of
variables one has indeed scaling at low binding for two-body states
with a good angular momentum $\ell$. The resulting plot for identified
halos will be shown later in figure \ref{fig:scalep2}.

It is less obvious how scaling appears in three-body systems. A more
thorough discussion of this can be found in \cite{Jen04}, it suffices
here to note that if a scaling radius $\rho_0$ can be found one can
introduce the dimensionless variables $\langle \rho^2 \rangle
/\rho_0^2$ and $mB\rho_0^2 /\hbar^2$. There are two slightly differing
choices of $\rho_0$, the first \cite{Rii00} proceeds in analogy to the
way the radius $\rho$ was introduced in equation (\ref{eq:rho}) and
makes use of two-body scaling radii $R_{ik}$ to define
\begin{equation}  \label{eq:rho1}
   m\rho_0^2 = \sum_{i<k} \frac{m_im_k}{M} R_{ik}^2  \;\;. 
\end{equation}
The second \cite{Jen03} builds on a deeper analysis of the
case where all potentials are square wells and defines 
\begin{equation}  \label{eq:rho2}
  \sqrt{m}\rho_0 = \frac{\sqrt{2}}{3} \sum_{i<k} \sqrt{\mu_{ik}} R_{ik}
\end{equation}
with $\mu_{ik}$ being
the reduced mass of the subsystem. If all masses and two-body scaling
radii are the same, both definitions reduce to $\rho_0 = R$.

\begin{figure}
  \includegraphics[height=9.5cm]{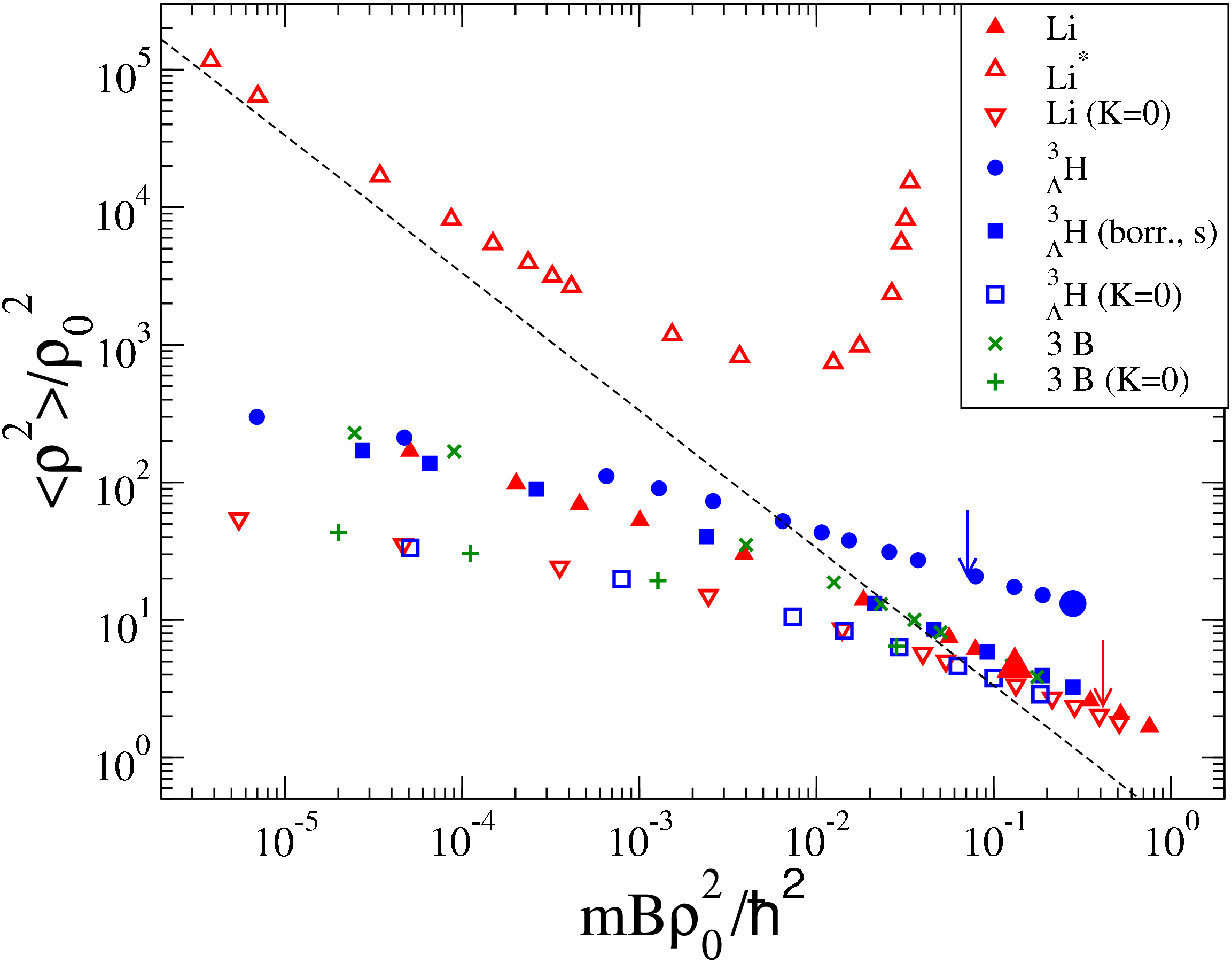}
  \caption{Three-body scaling plot for various systems as marked in the
    legend. The contribution to the mean square radius from the
    three-body radial coordinate is measured against the scaling
    radius from equation (\protect\ref{eq:rho2}) and displayed versus
    the scaled binding energy.
    The arrows indicate where two-body subsystems become bound
    and the dashed line (\dashed) indicates where Efimov states can
    appear. See the text for details. Courtesy E. Garrido.  \label{fig:scale3}}
\end{figure}

The main features of the classification of three-body systems
\cite{Jen03} are reproduced in figure \ref{fig:scale3} (the second
definition of $\rho_0$ is used here). Results are shown for three
model systems corresponding to $^{11}$Li, $^3_{\Lambda}$H and a
three-boson system where the interactions in all cases are varied to
give different binding energies, the large triangle and filled circle
corresponds to the physical $^{11}$Li and $^3_{\Lambda}$H. The
uncorrelated case with $K=0$ shows good scaling properties. When
correlations are allowed in the wavefunctions the main effect is to
increase the binding energy whereas the radii are less affected. The
resulting curves therefore lie to the right (or, equivalently, above)
the ones for $K=0$. The closest curve is the one for Borromean
systems. Here again a universal behaviour is seen for three different
systems. However, the physical hypertriton has a bound subsystem (the
deuteron) and is further away. That radii increase (for a given
three-body binding energy) as more and more subsystems become bound is
likely to be a general effect \cite{Fre06}. The arrows indicate the
positions where a two-body subsystem goes from being bound to
unbound. If the neutron-proton interaction is weakened the hypertriton
system (filled circles) eventually becomes Borromean and approaches
the other Borromean systems. The physical $^{11}$Li is of course
Borromean and when the neutron-$^9$Li interaction is increased nothing
spectacular happens to the ground state (filled triangles), but a very
extended excited state (open triangles) appears. This is a Efimov
state that appears along the dashed line just before the $^{10}$Li
subsystem becomes bound, increases in binding energy and finally
increases in size again when becoming unbound with respect to
n+$^{10}$Li once the latter has become bound.

The Efimov effect \cite{Efi70} appears when a two-body subsystem has a
very large scattering length $a$. This induces an effective potential
proportional to $\rho^{-2}$ in the three-body system for distances
between $R$ and $a$, a potential where successive excited states will
scale in energy as well as extension. More details can be found in
\cite{Nie01,Efi11}. These extraordinary states have been searched in
various systems, including nuclei and few-atom molecules, and were
finally observed indirectly in cold atom gasses a few years ago. Most
observations rely on the increase of the recombination rate of the
many-body system seen when the binding energy of the Efimov state goes
to zero, see \cite{Fer11} for a recent overview, but in a few cases
states have also been produced directly at finite binding energy
(\cite{Mac12} and references therein). So far experiments have probed
the binding energy systematics rather than the spatial extension of
the states. The one case where experimental information on sizes of
loosely bound molecules is available is the $^4$He dimer and trimer
where a very elegant experiment \cite{Bru05} have succeeded in
determining the sizes as 5.2(4) nm and 1.1(5) nm, respectively. An
excited He trimer Efimov state has been predicted and looked for for
several decades, but still not identified. In principle, much larger
molecular halo systems have already been produced via manipulation of
atomic scattering lengths in cold atom gasses and more cases are
expected to exist also as isolated systems.

To sum up, the only truly pathological behaviour occurs in two-body
systems where the scattering length $a$ can be arbitrarily large (and,
in the bound case, a halo state forms with arbitrarily small overlap
between the two bodies). A very large $a$ induces the Efimov effect
where the two-body correlations in a three-body system gives rise to
excited states with universal scaling properties and sizes reaching up
to $a$. In contrast to Efimov states that appear at the two-body
threshold, the Borromean three-body states that occur at the
three-body threshold are far more moderate in size for a given binding
energy. Three-body systems that appear above the Borromean curve in
the scaling plot therefore have substantial two-body correlations.

Going on now to $N$-body systems recent calculations for $N=4,5,6$
identical bosons \cite{Yam11} show that the systems remain finite even
for vanishing binding energy. The most important correlations remain
the two-body ones, even though higher-order correlations can also play
a role for excited states. The overall sizes seem to decrease as $N$
increases, but one can still observe systems that are significantly
larger than the scale set by $R$. Similar results are likely to be
present in the nuclear case and would be relevant for heavier nuclei
along the dripline than the ones accessed currently. If the nuclei
there are best described as $N$-body Borromean systems with $N$ larger
than three one would expect halo formation to be quenched. This
picture with several ``active'' neutrons around a core is probably the
reason, seen from few-body models, for the earlier mentioned decreased
size of halos when pairing mixes orbitals.

The hypertriton was used above to clarify the classification of
three-body systems and is best thought of as a $\Lambda$-halo around a
deuteron. Hypernuclear physics has probably more halos to offer
\cite{Hiy09}, but now with neutrons forming the halo rather than the
$\Lambda$-particle. Among the interesting systems are
$^6_{\Lambda}$He, where the outer neutron is expected to extend
further out than in $^6$He, and $^7_{\Lambda}$Be that with a binding
energy of around 0.3 MeV below the $^5_{\Lambda}$He+p+p threshold
probably is the best two-proton halo one could hope for. A recent
experiment \cite{Agn12} indicated that $^6_{\Lambda}$H is also bound
and quite close to the $^4_{\Lambda}$H+n+n threshold.

\subsection{Halo excitations}  \label{sec:excit}
The typical binding energy scale for nuclear halos is of order 100 keV
(less than about 1 MeV in light nuclei, decreasing with the mass
number as $A^{-2/3}$) and for molecular halos 1 $\mu$eV. There are
very few studies of the dynamics of molecular halos so the discussion
will from now on concentrate on nuclei.

States with halo structure can be excited in many different ways, the
ones of specific interest here are the ones that involve the halo
degree of freedoms. Since the halo extension depends so crucially on
binding energy one cannot expect the structure to be maintained under
e.g.\ rotation or isospin changes and we need to look carefully at
which states can be reached in excitations.

Strong interactions typically require more dedicated treatments,
whereas electroweak interactions can be treated perturbatively with
operators that may change spin and/or isospin, and may contain factors
$r^{\lambda}$ that enhance the tail. As will become obvious from the
following section halo excitations very often are dominated by
transitions to the continuum and also often dominated by the tail
properties. As also will be seen, in particular from the
electromagnetic probes in subsection \ref{sec:elmag}, these are not
qualitatively new features in nuclear physics but their magnitude puts
them on a quantitatively new level for halos.

Due to the importance of the continuum, the question of the continuum
structures seen in excitations must be considered carefully. A good
example is the E1 strength distribution of neutron halos that is now
understood as being single-particle strength going mainly directly to
the continuum \cite{Cat96} and not a semi-collective soft resonance,
see also \cite{Nak12}. As shown in detail in \cite{Myo98} for the case
of the $^{11}$Be E1 strength, one may choose to ascribe some of the
strength to resonances in the continuum. This particular work followed
the definition of Berggren \cite{Ber68} of a resonant state and his
elucidation of how one can go from a description entirely in terms of
a continuum to an equivalent description where resonances are included
explicitly. An important outcome of Berggren's analysis is that one in
general may expect a remaining non-resonant continuum contribution;
this is very much the case for the $^{11}$Be E1 strength and can be
expected also to be an important feature for other halo
excitations. The research theme of discrete states embedded in a
continuum is an important one in contemporary nuclear physics
\cite{Oko03} and will also be covered in these proceedings
\cite{Naz12}.

\section{Experimental probes of nuclear halos} \label{sec:probe} 
This section will present some of the different experimental ways of
testing and characterizing nuclear halo states. It is not possible to
make a complete coverage of this large field that also will be the
(partial) subject of quite a few of the other contributions to this
symposium. An early overview of the different types of probes of halos
was given in \cite{Rii92}. As remarked there one may classify the
probes according to how much they emphasize the tail, e.g.\ which
power of $r$ they correspond to. One should remember that the
probability of ``staying inside the binding potential'' remains finite
except for s-wave neutrons at vanishing energy, the large spatial
extent of halos is a result of a quite dilute very extended tail (cf.\ the
discussion after equation (\ref{eq:asymp})) and some probes will also be
quite dependent on core properties.

Combining information from different experiments gives a more accurate
picture of a given halo state. This section will focus on what
information may be obtained from the different experiments and whether
there are any generic effects or signals of the halo structure.  The
properties of some individual states will be summarized in section
\ref{sec:out}.

\subsection{Ground-state properties}
Due to the extreme sensitivity to binding energies, it is important to
know the mass of halo nuclei accurately. In most cases it is
sufficient that the binding energy is determined with an uncertainty
of 10 keV or less, but this is a rather challenging goal for nuclei
close to the driplines. General overviews of the techniques used in
mass measurements of radioactive nuclei can be found in
\cite{Lun03,Bla06}. The masses of halo nuclei were at first measured
via nuclear reactions or various time-of-flight measurements
\cite{Mit97}, but Penning traps have now also been brought to use
\cite{Dil12} and give an important step forward in accuracy as well as
precision. The lightest halo nuclei have now been measured with
sufficient accuracy and the challenges lie in producing neutron
dripline nuclei above Be in amounts that enable determination of the
masses of heavier halo candidates. A particular important
challenge is the mass of $^{19}$C that at the moment is deduced from
reaction measurement assuming it to be a good halo state rather than
from direct experiments.

Optical measurements on radioactive beams \cite{Che10} give access to
many nuclear groundstate properties, including charge radii, spin and
electromagnetic moments. These can give very precise information,
often on a level beyond what current nuclear theory can predict. The
charge radii are of course very valuable for a direct determination of
the sizes of proton halos, the experiments being sensitive to the mean
square radius of the total charge distribution. The best example of
this is the charge radius of $^{17}$Ne \cite{Gei08} that was found to
be clearly above the radius of heavier Ne-isotopes altough the
magnitude of the effect, an increase in rms radius less than 0.1 fm, is
clearly below what is found for good neutron halo nuclei. This is of
course consistent with the confining effect of the Coulomb barrier
that already in Ne is very noticable. Accurate measurements of the
charge radius of light neutron halo nuclei have also appeared during
the last few years \cite{Nor12}. Due to their high accuracy they are
important in giving constraints on core modifications, they can
furthermore for multi-neutron systems test the correlations in the
overall system through the sensitivity of the total charge radius to
the movement of the charged core around the centre-of-mass.

Magnetic dipole moments and electric quadrupole moments have also been
extracted for some halo nuclei (quite apart from the spin, an even
more basic property). The accurate measurements must be coupled to
model calculations to give the optimal information on the system, two
good examples are the magnetic moment of $^{11}$Be that is $\mu =
-1.6816(8) \mu_N$ \cite{Gei99} and the electric quadrupole moment of
$^{11}$Li whose ratio to that of $^9$Li, $|Q/Q_{core}| = 1.088(15)$
\cite{Neu08}, is similar to that of the rms charge radii for the two
nuclei. In general knowledge of moments of a halo candidate and the
``bare core'' nucleus can give important information on how inert the
core is in the halo state.

\subsection{Beta decay}
The halflife of a radioactive nucleus also belongs among its
groundstate properties, but will only in exceptional cases carry a
clear signature of a halo structure. Beta decay probabilities depend
on the overlap between initial and final state wavefunction and the
effects of a halo on an overlap are in most cases below a factor of
two which means one needs a good understanding of the core structure
of the initial and final state in order to draw conclusions. However,
beta decay brings information on halos in other ways (see
\cite{Rii06,Nil00,Pfu12} for more detailed reviews).

First, beta decay may help pin-pointing the exact configuration of a
halo. A good example is $^{11}$Li where already the total halflife
indicates that the two last neutrons must reside partly in the
sd-shell rather than only in the p-shell \cite{Bar77}. Individual
decay branches of course can be used to strengthen this argumentation
\cite{Nil00}. Secondly, there are indications that the halo and core
beta decays decouple at least in some cases (it is not yet clear
whether this will be a general feature for good halo states). The
prime example of a core decay that reappears in the decay of the halo
nucleus is that of $^{12}$Be where more than 99\% of the decay goes to
the $^{12}$B $1^+$  ground state and one correspondingly finds that
most of the $^{14}$Be decays goes to a slightly unbound $1^+$ state in
$^{14}$B \cite{Jon04,Nil00,Tim96}.

An even more convincing case would be a decoupled halo decay. This
appears to occur at least for two-neutron halo systems as beta-delayed
deuteron emission. The current understanding of this decay mode is
that it proceeds directly to the continuum \cite{Nil00,Pfu12}, a decay
of the two halo neutrons directly to a deuteron in the periphery of
the total system. Two cases are established so far, for $^6$He the
intensity is low due to a cancellation of contributions from inner and
outer parts of the wavefunction whereas $^{11}$Li, where the energy
spectrum was recently measured \cite{Raa08}, offers the possibility of
a more stringent test once the final state interaction between the
deuteron and the $^9$Li core is known better. It will be interesting
to see whether other halo nuclei will give similar unique decays. At
least for the two known cases it is clear that the contribution from
the non-resonant continuum is essential for describing the decays.

\subsection{Electromagnetic processes}  \label{sec:elmag}
Electromagnetic transitions of type E$(\lambda)$ or M$(\lambda+1)$
have operators that contain a factor $r^{\lambda}$ which enhances the
tail behaviour and makes these processes very sensitive to halo
formation. It is worth noting that halos in excited states can be
probed as well provided one can single out transitions
that involve these states. The emphasis on the wavefunctions at
large radii was known also before reaction experiments indicated the
existence of halo states, and several earlier papers identified the
main features of what we now recognize as being halo signatures.

For transitions between bound states the classic case is that of
$^{11}$Be where the E1 transition between the first excited 1/2$^-$
state and the 1/2$^+$ ground state could only be reproduced if the
spatial extent of the wavefunctions was taken into account
\cite{Mil83}. Transitions between a bound state and the continuum were
known to be important for the deuteron, but the case of proton
radiative capture into a spatially extended state was also clearly
understood for $^8$B and the first excited state in $^{17}$F
\cite{Chr61,Rol73}. In these cases direct capture, i.e.\ contributions
from the non-resonant continuum, are more important than resonance
capture in line with the discussion in section
\ref{sec:excit}. The proton radiative capture was investigated early
on due to its importance for nuclear astrophysics, but a similar
sensitivity can be expected in neutron radiative capture \cite{Rii92}
and should again give a low lying non-resonant peak \cite{Ots94}.

Electromagnetic dissociation is the inverse reaction to radiative
capture and will therefore be as sensitive to halo structures.  It is
now recognized as a key probe for nuclear halo states and an important
dynamical consequence of the special structure, as first pointed out in
\cite{Han87} and observed experimentally shortly after \cite{Kob89},
see also \cite{Nak12}.  The prediction in \cite{Han87} was based on
sum-rules for the E1 strength. These are easily generalized and give
for a two-cluster model \cite{Sag92}, where the clusters have charge
and mass number $Z_i$ and $A_i$ and distance $r$ between them, a
contribution to the energy-weighted sum of
\begin{equation}
   \frac{9}{4\pi} \frac{(Z_1A_2-Z_2A_1)^2}{(A_1+A_2)A_1A_2} \frac{\hbar^2e^2}{2m}
\end{equation}
and for the non-energy-weighted sum
\begin{equation}
  \frac{3}{4\pi} \frac{(Z_1A_2-Z_2A_1)^2}{(A_1+A_2)^2} \langle r^2 \rangle e^2 \;.
\end {equation}
Compared to a normal state, a halo state will therefore give more
strength that appears at lower excitation energy than usual and if the
specific contribution to the E1-strength from the halo degree of
freedom can be extracted one can derive its spatial extent. The
E1-strength will carry the dominant signal for neutron halos, whereas higher
orders will also be affected for proton halos.
Concerning the strength distribution, the simple structure of halos
implies that simple analytic models \cite{Typ04,Nag05,Typ05} can be
expected to give the main features. Simple analytic expressions for
the photodissociation cross-sections valid for loosely bound nuclei
have also been derived \cite{Kal96}.

As discussed above, and in complete correspondence to radiative
capture, the dissociation reactions will involve mainly the
non-resonant continuum. Their importance will be seen clearly in the
next section, one good example is the identification of $^{31}$Ne as a
halo candidate based on its large one-neutron removal cross-section
\cite{Nak09}.

\subsection{Nuclear reactions}
Most studies on halos have been made with nuclear reactions. I can
here only give a brief overview of this large field that also is
covered in several other contributions in these proceedings, in 
particular \cite{Bon12}. Detailed reviews of the reaction theory of
halo nuclei can be found in
\cite{Ber93,Jen04,Can06,Kee07a,Kee09}. 

It is customary to classify the reactions according to the beam
energy, low energy reactions taking place around the Coulomb barrier,
intermediate energy reactions around the Fermi energy,
and above this high energy reactions (extending to
relativistic energies) where the
smaller nucleon-nucleon cross-section and the shorter interaction
times makes reaction mechanisms simpler. The reactions mechanisms do
evolve gradually as the energy increases so a strict division between
the three regions is not possible, but rough dividing points are
somewhat above 10 MeV/u and around 100 MeV/u.

As realized essentially from the start the ``halo-removal'' channel
(one-neutron removal for one-neutron halos, two-neutron removal for
two-neutron halos etc) will, for both strong and Coulomb interactions,
have a significant cross-section and will furthermore be clearly
influenced by the halo structure. The main focus has been on this
channel, this is fully justified but has implied that possible
information from other channels often has been neglected.

\subsubsection{Interaction cross-section}
The experiments that triggered the interest in halos
\cite{Tan85a,Tan85} measured the total interaction cross-section for
different isotopes and observed a clear enhancement for halo
nuclei. The key
point is that the strong interactions between nucleons makes reactions
so likely that nucleons ``shadow'' each other so that compact systems
with more shadowing will have reduced cross-sections. (This is in
contrast to electromagnetic or weak interactions that only observe the
total ``charge'' of a nucleus.)
As gradually became clear it is important to have
a correct understanding of the reaction mechanism in
order to extract reliable radii for the halos \cite{AlK96}. This is
now the case for reactions at high energy, a compilation of matter
radii extracted from such experiments can be found in
\cite{Oza01}. There are many attempts at extending the theory also to
intermediate and low energies, but for the moment extracted values
from these energy ranges should be treated with more
caution. Nevertheless, the large deduced radius for $^{22}$C
\cite{Tan10} clearly identifies this nucleus as a new halo candidate
even though the uncertainty of its radius is considerable.

For good halo systems it was
pointed out in \cite{Yab92} that the ``halo removal'' cross-section at
high energy will
be equal to the difference in reaction cross-sections for the halo
nucleus and the core nucleus, a quite direct reflection of the
decoupling of the halo system into halo and core parts.
In line with this, the charge-changing cross-sections for halo nuclei
and their core have been found to be essentially the same \cite{Bla92}.

\subsubsection{High to intermediate energy probes}
Many types of experiments have been developed for use at high and
intermediate energy. Among the first were measurements of the
transverse momentum distributions \cite{Kob88,Ann90} of reaction
products from break-up reactions. Shortly after longitudinal momentum
distributions of the core fragment \cite{Orr92} became available,
these are less affected by the reaction and --- if the reaction
mechanism is disregarded --- can be easily interpreted (a large
spatial extension corresponding to a narrow momentum width).
Surveys of longitudinal momentum distributions have been carried out for
many neutron-rich nuclei \cite{Sau00,Rod10}.
The reaction mechanism of course needs to be taken into account as
well in order to extract quantitative information. On the experimental
side an often needed refinement is a more exclusive measurement where
the final state of the core is detected (of course only relevant when
the core has several particle bound levels). A gating on specific
final states is needed in order to get a clean
distributions as first shown for $^{11}$Be \cite{Aum00}.

Most recent work has focussed on core longitudinal momentum
distributions in the halo-removal channel.
Although not as widely appreciated, it is
also valuable to look at neutron momentum distributions at high
reaction energy for processes where the core is removed by break-up
\cite{Nil95} or kicked out by a Coulomb field (this should e.g.\ give
a very clear signature in the neutron-neutron correlation
\cite{Gar01}).

Experimental set-ups are now available at most facilities to allow
complete kinematics experiments to be carried out. Here all outgoing
fragments from the break-up are recorded combined with gamma-ray
detection at the taget position, in several cases with proton targets
the recoiling proton has also been detected. This allows
reconstruction of the excitation energy in the final state as well as of
various distributions in sub-systems, several examples will be given
in the next section. This of course puts more stringent demands on the
reaction models, see \cite{Jen04,Can06} for the status for high-energy
experiments and \cite{Cap12} for a recent comparison of diferent
breakup models at intermediate energy.

A particular example is that of elastic scattering at high energy on a
proton target. This well established method has been employed in
several experiments for the lightest halo nuclei and gives a quite
accurate description of the density profile of the nuclei allowing for
an independent determination of the matter radii.  A recent example is
the work on the heavy Be isotopes leading up to a determination of the
radius of $^{14}$Be \cite{Ili12}.

\subsubsection{Intermediate to low energy probes}
The, already established, many different types of lower energy
reactions have essentially all been applied to halo nuclei once they
became available as secondary beams at lower energy. This includes in
particular elastic scattering, fusion reactions, transfer reactions
and break-up reactions, see \cite{Can06,Kee07a,Kee09} for
reviews. This field is still evolving and is still to some extent
focussed on identification of halo signatures rather than extraction
of halo properties. I will give just two different examples of this.

In elastic scattering at low energy the nucleus has time to adapt
during the collision giving e.g.\ unique polarization effects that do
not appear at higher energy.  One example is the elastic scattering of
$^{11}$Be on Zn \cite{DiP10} where careful comparison with the also
measured elastic scattering of $^{9,10}$Be gave a clear halo signature.

Transfer reactions are a powerful tool for nuclear structure studies
and secondary beam techniques have evolved so that it now has been
possible to do e.g.\ (p,t) reactions both on $^8$He \cite{Kee07} and
$^{11}$Li \cite{Tan08}.  The basic theoretical procedures are still
being critically discussed, see e.g.\ \cite{Das09}. One recent
interesting suggestion is to focus on the ratio of angular
distributions for elastic breakup and scattering \cite{Cap11} that
appears less sensitive to the reaction mechanism giving more direct
access to the halo properties.

\section{Knowns and unknowns of nuclear halo states}  \label{sec:out}
The very first experiments on nuclear halo states gave only the
rudimentary properties of the states, but the developments in
experimental techniques (both in production and in detection methods)
have given an increasingly more detailed characterization of the
lightest states such as $^{11}$Li and $^6$He. First observations of
halo states typically take place at in-flight facilities, but
ISOL-based facilities play an important role in refining our
understanding in the later stages of experiments.  The new generation
of radioactive beam facilities \cite{Blu12}, starting with RIBF in
RIKEN, will allow studies to continue in new mass regions; several
promising halo candidates have already been identified recently.

This section will first give an overview of what is presently known on
nuclear halo states with a focus on the most pronounced cases. The
discussion will hopefully indicate what is needed in order to
establish future halos. It is very encouraging that the field has
reached a stage where one can cross-check results, e.g.\ on matter
radii, by comparing results obtained with different methods. 
Nevertheless, there are still several unknowns. The last subsection
gives a list (with a personal bias) on what the remaining open
questions are in the field.

\subsection{Established halo states}
\begin{figure}
  \includegraphics[height=10.0cm]{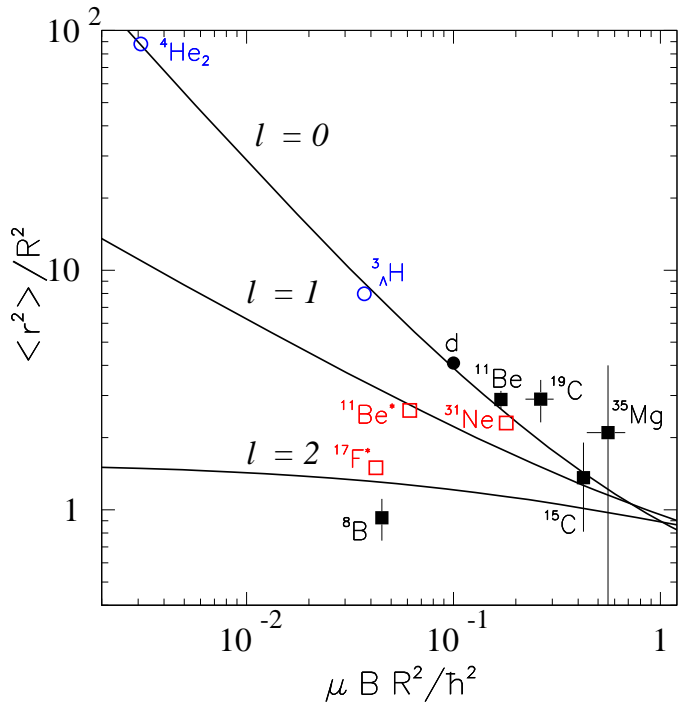}
  \caption{Scaling plot for two-body halo systems. The filled circle
    denote the deuteron, the filled squares nuclei where radii were
    extracted from experimental interaction cross-sections, the open
    squares are simple model estimates and the open circles
    theoretical calculations. See the text for details.
    \label{fig:scalep2}}
\end{figure}

\begin{figure}
  \includegraphics[height=10.0cm]{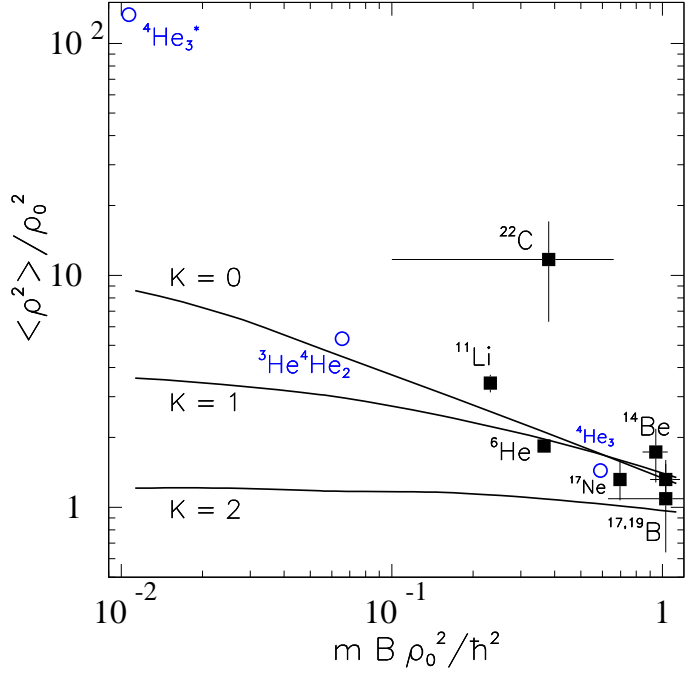}
  \caption{Scaling plot for three-body halo systems. The filled
    squares denote nuclei where radii were extracted from experimental
    interaction cross-sections and the open circles results of
    theoretical calculations. See the text for details.
    \label{fig:scalep3}}
\end{figure}

In order to use a few-body picture for the halos the amount of
``configuration mixing'' should be limited, i.e.\ there should in the
nuclear case be mainly one component in the wavefunction with one or
two nucleons around a core (that often, but not necessarily, will be
in the ground state). This appears to be fulfilled for most of the
nuclei considered below.
Scaling plots updated with the latest experimental information are
shown as figures \ref{fig:scalep2} and
\ref{fig:scalep3}. For the three-body halos equation (\ref{eq:rho1})
is used to define $\rho_0$, the alternative definition in equation
(\ref{eq:rho2}) gives a $\rho_o^2$ that for the nuclear two-neutron
halos is smaller by about a factor 1.7.
It may be of interest to note that systems where the probability of
being outside of the potential range is higher than 50\% corresponds
to a value just below 2 for the scaled square radii. As can
be seen in the figures this restricts the number of pronounced halos.

For easier reference some properties of selected halo systems are
collected in tables \ref{tab:2body} and \ref{tab:3body}. The tables
list the root mean square radii for the total system as well as for
the core nucleus, the components of the system and the binding energy
\cite{Aud11} for the halo particle(s). 
Except where discussed explicitly in the following the
experimental values for radii are taken from \cite{AlK96a,Oza01}, in
cases where several analyses have been made the ``few-body'' results
(that take correlations in the system into account) are used.
For two-body halos the angular momentum of the halo particle is given,
for three-body halos the possible angular momenta for a halo neutron with
respect to the core is given. The deduced mean square radius of the
halo is listed at the end. Since selections between different data
had to be done in several cases the following detailed discussion
should be consulted to get a complete overview for a specific
state. The assumptions made are also discussed there.

\begin{table}
  \caption{\label{tab:2body}Properties of some two-body halos.}
\begin{tabular}{@{}lccccccc}   \br
      System & Composition & $\ell$ & $B$ (MeV) & $r^{rms}_{core}$ (fm) &
      $r^{rms}_{tot}$ (fm) & $\langle r^2 \rangle$ (fm$^2$) & $\langle
      r^2 \rangle /R^2$ \\   \mr
     d & n+p & 0 & 2.225 & --- & 1.9754(9) & 15.61(2) & 4.2 \\
     $^8$B & $^7$Be+p & 1 & 0.136(1) & 2.31(5) & 2.50(4) & 14(3) & 0.93(18) \\
     $^{11}$Be & $^{10}$Be+n & 0 & 0.502 & 2.28(2) & 2.90(5) & 44(4) & 2.9(3)  \\
     $^{11}$Be$^*$ & $^{10}$Be+n & 1 & 0.182 & 2.28 & --- & 40 & 2.6 \\
     $^{15}$C & $^{14}$C+n & 0 & 1.218(1) & 2.30(7) & 2.50(8) & 21(8) & 1.4(5) \\
     $^{17}$F$^*$ & $^{16}$O+p & 0 & 0.105 & 2.72 & --- & 28 & 1.5 \\ 
     $^{19}$C & $^{18}$C+n & 0 & 0.58(9) & 2.82(4) & 3.23(8) & 58(11) & 2.9(6) \\ 
     $^{31}$Ne & $^{30}$Ne+n & 1 & 0.3(2)$^{a}$ &  ? & ? & 60$^{a}$ & 2.3$^{a}$ \\ 
     $^{35}$Mg & $^{34}$Mg+n & ? & 1.0(2) & 3.23(13) & 3.40(24) & 50 & 2(2) \\ \br
    \end{tabular}

    $^{a}$ See the text.
\end{table}

\begin{table}
  \caption{\label{tab:3body}Properties of some three-body halos.}
\begin{tabular}{@{}lccccccc}   \br
      System & Composition & $\ell$ & $B$ (MeV) & $r^{rms}_{core}$ (fm) &
      $r^{rms}_{tot}$ (fm) & $\langle \rho^2 \rangle$ (fm$^2$) & 
      $\langle \rho^2 \rangle /\rho_0^2$ \\   \mr
    $^6$He & $^4$He+n+n & 1 & 0.975 & 1.58(4) & 2.54(4) &
    28.6(1.3) & 1.84(10) \\
     $^{11}$Li & $^9$Li+n+n & 0, 1 & 0.369(1) & 2.30(2) & 3.53(10) &
     89(8)  & 3.4(3) \\
     $^{14}$Be & $^{12}$Be+n+n & 0, 1, 2 & 1.26(13) & 2.59(6) & 3.10(15)
     & 54(13) & 1.7(4) \\
     $^{17}$B & $^{15}$B+n+n & 0, 2 & 1.34(17) & 2.59(3) & 2.90(6) &
     42(6) & 1.3(2) \\   
     $^{19}$B & $^{17}$B+n+n & 2 ? & 1.0(4) & 2.90(6) & 3.11(13) &
     41(16) & 1.1(5) \\  
     $^{17}$Ne & $^{15}$O+p+p & 0, 2 & 0.933 & 2.44(4) & 2.75(7)
     & 39(7) & 1.3(3) \\
     $^{22}$C & $^{20}$C+n+n & 0 ? & 0.4(3)$^{a}$ & 2.98(5) & 5.4(9)
     & 460(210) & 12(5) \\ \br
    \end{tabular}

    $^{a}$ See the text.
\end{table}

\subsubsection{The deuteron}
The deuteron is obviously clustered into a proton and a neutron and in
this sense has no core. Historically it is, as remarked in \cite{Han95},
``the forerunner of all halo states'' and where the main reaction
mechanisms for halos were discovered. (In the words of Gregers Hansen
\cite{Ann94} it is ``the mother of all halo nuclei and certainly the
only one with both a proton and a neutron halo''.)  The rms radius has
been measured accurately from optical measurements \cite{Hub98} and is
1.97535(85) fm. Its exact position in figure \ref{fig:scalep2} depends
on the choice of $R$ which is of order 1.9--2.0 fm. Recent
reaction work on deuterons has focussed on relativistic scattering
experiments, but a review of low-energy deuteron break-up can be found
in \cite{Bau76}.

\subsubsection{$^{11}$Li}
This was among the first identified halo states \cite{Tan85} and is still the
archetype of a two-neutron halo. Much of the existing data have been
summarized and compared to a three-body model in \cite{Shu09}. The
following brief discussion mainly illustrates the large amount of data
available on $^{11}$Li.

The total matter radius is extrated \cite{AlK96a} as 3.53(10) fm from
the interaction cross-section and 3.71(20) fm from high-energy elastic
scattering \cite{Dob06}. The two radii determinations are consistent,
but since the radii of $^9$Li also differ slightly from the two
methods the deduced values of $\langle \rho^2 \rangle /\rho_0^2$ are
essentially identical: 3.4(3) and 3.5(6), respectively. The former
one is the value included in table \ref{tab:3body}.

The magnetic dipole moment is slightly larger than for the core $^9$Li
(closer to the single-particle value), as mentioned above this also is
the case for the electric quadrupole moment. The charge radius data
may indicate the need to include also core excitations \cite{San06}, a
more extensive discussion is given in \cite{Nor11}. The charge radius
is bascially consistent with the radii determined from reaction
experiments, but the simple picture of an inert $^9$Li core and two
halo neutrons coupled to angular momentum zero must be modified
slightly to make all data fit \cite{Shu09,Nor11}.

A naive filling of orbits from a simple shell model scheme would have
placed the two halo neutrons in p-waves. The need for s-wave
components was deduced already from the beta-decay halflife
\cite{Bar77}, but also arises from an analysis of the reaction
experiments \cite{Tho94}. Rather direct evidence for different parity
states was obtained from the observation of anisotropic
angular distribution between fragments \cite{Sim99}. The s- and p-wave
components have roughly equal weight, but smaller contributions of
e.g.\ d-wave could also be present.

In the halo break-up channel $^9$Li has a longitudinal momentum
distribution with FWHM of order 40--50 MeV/c depending on target and
beam energy \cite{Orr97}. An analysis of two-neutron interferometry
\cite{Mar00} gave an average distance between the two neutrons of
6.6(1.5) fm.  The E1 strength distribution measured in detail
\cite{Nak06} led to a deduced value for the root-mean-square distance
between the core and the centre-of-mass of the two halo neutrons of
5.0(3) fm. All in all the combined data indicate that the two halo
neutrons tend to be on the same side of the core, an asymmetry made
possible by the configuration mixing of s- and p-waves.

Through a $^{11}$Li(p,n)$^{11}$Be$^*$ reaction af 64 MeV/u the
isobaric analogue state of $^{11}$Li was observed to have a smaller
Coulomb displacement energy than for other Li isotopes \cite{Ter97},
consistent with calculations including the halo \cite{Suz98}. See also
\cite{Suz91} for a general discussion of the relevant isospin multiplets.

\subsubsection{$^{6,8}$He}
The nucleus $^6$He was also among the first halo states to be
identified \cite{Tan85a}. Its structure is well understood with the
two neutrons being in p-orbits outside the alpha particle core. It is
often used for detailed checks of our understanding of reaction
mechanisms and other effects. It has been studied extensively in
complete kinematics experiments at GSI and very complete data on its
behaviour in reactions is now available \cite{Chu05}.

Adding two neutrons to $^6$He gives $^8$He that is considerably more
bound. It appears to be better thought of as four neutrons outside an
alpha particle core rather than two neutrons outside $^6$He. Its
spatial extension is not as large and it is probably better to
consider it a skin nucleus \cite{Tan92}.

Also here there is agreement between total matter radii extracted from
interaction cross-sections and high-energy elastic scattering. The
matter radius of $^8$He is only slightly larger than the one of $^6$He
and the charge radius is actually smaller, this is understood in terms of
the correlations between the halo neutrons in $^6$He \cite{Mue07}.

For both $^6$He and $^8$He many experiments have looked at (in)elastic
scattering, fusion reaction and transfer reactions.

\subsubsection{$^{11}$Be}
Following the deutron this is the first and most studied one-neutron
halo. A particular feature is that the first excited state also is a
halo, but with the neutron in a p-wave rather than an s-wave (a simple
estimate for the radius of the excited state is given in table
\ref{tab:2body}). This special structure is reflected in the strong E1
transition between the two states \cite{Mil83}.

The matter radius is extracted as 2.90(5) fm from interaction
cross-sections \cite{AlK96a} or as 2.91(5) fm with the alternative
analysis in \cite{Oza01}. The deduced rms value for the distance
between the core and halo neutron is then 6.6(3) fm, which can be
compared with 6.4(7) fm extracted from Coulomb dissociation at 72
MeV/u \cite{Nak94}. Later determinations from other Coulomb
dissociation experiments tend to give lower values for the distance
\cite{Pal03,Fuk04}, but the charge radius measurement rather indicates
a distance of 7.0 fm \cite{Nor09}.

The longitudinal momentum of $^{10}$Be in the one-neutron removal
channel has a FWHM around 45 MeV/c \cite{Orr97}. There are several
excited states in $^{10}$Be and by gating on observed gamma rays
detailed studies have been made of the one-neutron removal reaction
leading to specific final states \cite{Aum00,Pal03}.

The core is not completely inert and the wavefunction will contain a
component with a $2^+$ excited core. Several experiments have
addressed this issue and by now a consistent understanding of the size
of this effect has been obtained \cite{Sch12}, about 70\% of the
wavefunction will be the ``classical'' one with a $^{10}$Be ground
state.

\subsubsection{$^8$B}
The large spatial extension of the wavefunction of the p-wave
proton in the ground state of $^8$B was, as mentioned above in section
\ref{sec:elmag}, already noted early in connection with studies of
proton radiative capture \cite{Chr61}. The $^7$Be(p,$\gamma$)$^8$B
remains astrophysically interesting which has given rise to much
experimental activity on this nucleus.

Due to the combined effect of the Coulomb and centrifugal barrier its
radius is not as pronounced as for $^{11}$Be even though the binding
energy is low. Several detailed reaction studies have been carried out,
e.g.\ a complete kinematics experiment \cite{Cor03} that determine the
core excited state component to be 13(2)\%. One can also mention the
measurement \cite{Sum06} of the electric quadrupole moment of
$+64.5(1.4)$ mb, a value that seems to be sensitive to the extended
proton tail.

\subsubsection{$^{17}$F$^*$}
The very large spatial extension of the wavefunction of the s-wave
proton in the first excited $1/2^+$ state in $^{17}$F was, as for
$^8$B, noted from proton radiative capture, see in particular
\cite{Rol73} for a very clear exposition and \cite{Ben00} for a later
calculation within the continuum shell model.  The recognition that
this corresponded to a halo state \cite{Rii92} prompted investigations
of possible effects in beta decay into this level. A clear deviation
of mirror symmetry was seen, but the theoretical interpretation also
depends on the description of the mother nucleus, $^{17}$Ne, and gives
only indirect information on the state, see the review in
\cite{Nil00}. The radius quoted in table \ref{tab:2body} is a simple
model estimate. The special status of this state was apparently
rediscovered in \cite{Mor97}.

\subsubsection{$^{17}$Ne}
This two-proton halo candidate most likely has a well-defined core in
$^{15}$O. The beta decay data may indicate changes of orbit occupation
compared to the mirror decay of $^{17}$N, see the discusison in
\cite{Nil00} and the detailed calculation within the shell model
embedded in the continuum approach \cite{Mic02}.

As already mentioned its charge radius has been measured \cite{Gei08}
and is in agreement with the matter radius of 2.75(7) fm extracted
from the interaction cross-section \cite{Oza01}. This is probably the
best candidate for a two-proton halo in normal nuclei, but has a
moderate spatial extension.

\subsubsection{$^{14}$Be}
The binding energy of this two-neutron halo candidate is above 1 MeV. 
Its matter radius has been extracted as 3.10(15) fm from interaction
cross-sections \cite{Oza01} and 3.25(11) fm from high-energy
elastic scattering \cite{Ili12}, again in good agreement.

It seems increasingly likely that this system has parallels to $^8$He
in the sense that $^{12}$Be cannot be considered an inert core. No
experiment has so far reported on core-excitations in two-neutron
removal reactions, but $^{12}$Be is known to have bound $2^+$, $0^+$
and $1^-$ states and in particular the excited state $0^+$ state is
very interesting in this respect. More experimental information is
needed to clarify our understanding of this interesting system in
between normal nuclei and the pronounced two-neutron halos.

\subsubsection{$^{15}$C}
The one-neutron binding energy is here just above 1 MeV, but the last
neutron seems to be a good single-particle s-state and the system is
therefore a good example of a state in between normal nuclei and well
developed one-neutron halos. This is reflected both in its
interaction cross-section and in the width of the longitudinal
momentum distribution after one-neutron removal, note that this
nucleus and $^{22}$N appear to have a quite similar behaviour in the
general surveys \cite{Sau00,Rod10}.

\subsubsection{$^{19}$C}
The first indications of a halo in this nucleus came from the
longitudinal momentum distribution after one-neutron removal with a
FWHM of only 44(6) MeV/c \cite{Baz95}. The direct measurements of the
binding energy have too low precision and the currently used value of
0.58(9) MeV is derived e.g.\ from analysis of Coulomb dissociation in the
one-neutron removal channel assuming the state to be a good halo
\cite{Nak99}. A direct precise measurement of the binding energy would
be very helpful.

All reaction data are consistent with this system being a good
single-particle halo. The observation of two gamma rays \cite{Ele05}
from excited states in $^{19}$C indicates an interesting low-lying
structure (compatible with expectations from shell models) and
incidentally puts a lower limit of the binding energy.

\subsubsection{$^{17,19}$B}
The matter radii of the two heaviest boron isotopes was measured in
\cite{Suz99}. A few more reaction studies have been made on $^{17}$B
(e.g.\ demonstrating the presence of a bound excited state just below
1.1 MeV), but very little is currently known about $^{19}$B. Two
different values have been extracted for the matter radius of
$^{17}$B, namely 2.90(6) fm from the optical limit and 2.99(9) fm from
a few-body approach \cite{Oza01}. The former one is used in table
\ref{tab:3body}, if the latter is used the value of $\langle \rho^2
\rangle /\rho_0^2$ will increase to 1.6(3). It could well be that
$^{17}$B is a more extended system than $^{19}$B in spite of being
more bound. However, one should note that is was suggested
\cite{Suz99} that $^{19}$B is better though of as having four valence
neutrons around a core so the values derived from a three-body model
could well be inapplicable. More data are clearly needed on $^{19}$B,
but it is striking that we in both $^8$He and
$^{19}$B, the two nuclei investigated so far where both the one- and
three-neutron removal leads to an unbound system (five-body Borromean
nuclei), see indications for a four-neutron structure rather than a
two-neutron structure build upon the intermediate nucleus. (A similar
situation could actually also be present in $^{14}$Be.) This is of
course relevant for the extrapolation to much heavier neutron dripline
nuclei where five-body Borromean (and seven-body etc.) nuclei will be
very common.

\subsubsection{$^{22}$C}
This $N=16$ nucleus is a quite interesting system. The radius
of 5.4(9) fm extracted from intermediate energy reaction cross-sections
\cite{Tan10} is very large and would make this by far the
largest nuclear halo known to date, see table \ref{tab:3body}. However, it
seems likely that the 5.4 fm is an overestimate (note that the error is quite
significant), since momentum distributions from two-neutron removal on
a carbon target \cite{Kob11} do not show the same extreme
result. Future experiments should be able to sort this out: a radius
above 5 fm should give rise to a very large Coulomb dissociation
cross-section unless there happens to be a strong anti-correlation
between the two halo neutrons (so that the core and centre-of-mass of
$^{22}$C almost coincide). On the other hand, should this large radius
turn out to be correct it would indicate that there is a very strong
core-neutron substructure, cf.\ the discussion around figure
\ref{fig:scale3}. The currently known binding energy is much too
uncertain and the value used in table \ref{tab:3body} and figure
\ref{fig:scalep3} is a guess reflecting the prejudice that the true
value must be below 1 MeV.

\subsubsection{$^{31}$Ne}
This nucleus was first observed at RIKEN \cite{Sak96} and it is also
there that one recently, after the start of RIBF, has been able to
determine its Coulomb cross-section \cite{Nak09} and interaction
cross-section \cite{Tak11}. Both show the typical enhancement for a
good halo state. Most properties of this nucleus have still not been
measured with sufficient accuracy, but it is believed (from
systematics and the analysis of the Coulomb break-up) that the halo
neutron is in a p-wave.  Several theoretical calculations on this
interesting system are already available.

Its binding energy has an uncertainty of more than one MeV, but it
should be within a few hundred keV of the threshold to fit the
available reaction data, so I have arbitrarily put it at 0.3(2)
MeV. The order of magnitude of the one-neutron Coulomb dissociation
cross-section indicates a halo radius of the order of the one in
$^{19}$C and a scaled extension that is slightly lower (due to the
larger core nucleus). I have used this for the rough estimates in
table \ref{tab:2body}.

\subsubsection{Heavier nuclear states}
There is by now sufficient experimental information available to
conclude that we have identified all possible pronounced ground state
halos up to about mass 30. Above this mass we do not expect proton
halos to be very extended, and the current information on the neutron
dripline candidates is quite limited. However, a recent reaction
experiment \cite{Kan11} has pointed to $^{35}$Mg (extracted radius of
3.40(24) fm with a core radius of 3.23(13) fm) as a possible
interesting nucleus. The precision must be improved in order to make a
definite conclusion based on interaction cross-sections only.

The effect of the halo neutron(s) becomes less and less on a relative
scale as we go to higher masses which explains the higher requirement
for the precision on the cross-sections. It may be easier to detect
halo states in the halo-removal channel via the Coulomb dissociation, see
section \ref{sec:elmag} above and \cite{Nak12}. Some of the
interesting nuclei that could be studied in the future in this way are
the heavy Mg and Si isotopes.

\subsubsection{Halos in excited nuclear states}
It is more difficult to extract information on halos in excited
states, but as mentioned above it is possible in electromagnetic
transitions (going to or from the state). The other possibility is to
look for signals in nuclear reactions that populate the
state. Recently, it was suggested \cite{Ogl11} that a general signal
may be an enhanced diffraction radius in intermediate energy
scattering. Transfer reactions into a candidate state can of course
also provide information, if sufficient knowledge is available on the
reaction partners.

Apart from the excited $1/2^-$ state in $^{11}$Be and the $1/2^+$
state in $^{17}$F mentioned earlier other interesting halo candidates
are the $1^-$, $2^-$ states in $^{10}$Be \cite{AlK06} if their
configurtion is an s-wave neutron around a $^9$Be core. Two-neutron
halos in excited states may also exist e.g.\ in $^{12}$Be
\cite{Rom08}, but for most of these candidates there is insufficient
experimental evidence. More candidates are listed in tables I and II
in \cite{Jen04}.

One special state where initial experiments have been performed is the
$0^+, T=1$ excited state in $^6$Li \cite{Ara95} that should have a
neutron-proton halo and has been probed via pionic fusion \cite{And00}
as well as the $^1$H($^6$He,$^6$Li)n reaction \cite{LiZ02}. It is the
isobaric analogue state of the ground state of $^6$He.

\subsubsection{Halos in other systems}
Theoretical predictions for hypernuclear halos have also been
performed in several cases. Predictions for the hypertriton \cite{Cob97}
are included in figure \ref{fig:scalep2}. The experimental
information is very hard to obtain and at the moment only binding
energies are known, with considerable uncertainties, and nothing has
been established concerning the spatial extensions.

Theoretical calculations for the $^4$He dimer and three trimer states
\cite{Nie98} are also included in the scaling figures. The other
few-atom molecular states that appear in current cold atom gas
experiments can have even larger sizes that furthermore can be tuned
via an external applied magnetic field. They have not been isolated
as single molecules yet, but would in the scaling plots occur along the
$\ell=0$ line in figure \ref{fig:scalep2} and along the dashed
diagonal in figure \ref{fig:scale3}.

\subsection{Open questions}

Halo physics has made impressive progress during the 25 years since
they were discovered. There are nevertheless still
several open questions we need to address. The following incomplete
list are the main ones coming to my mind when preparing this
contribution.
\begin{itemize}
\item Where will nuclear halos heavier than the currently known occur
  in the nuclear chart ? Somewhat coupled to this is the question of how much
  single particle character remains in heavier nuclei.
\item Are there dynamical characteristics of \emph{skin} nuclei ?
\item Does closeness to continuum promote (or stabilize) halo cluster
  structures ? For the general case of clusters this may be the case
  \cite{Oko12}, recent work \cite{Zin08,Ebr12,Ebr12a} is starting to
  establish quantitative criteria for clustering. The finding so far
  that most halo candidate states (with low angular momenta) end up
  having a good halo structure may be a ``selection bias'' since all
  cases are among the light nuclei, but the continuum could also play a
  more active role.
\item When is the concept of a resonant continuum useful ? As seen
  above the concept of a non-resonant continuum is essential in order
  to understand all aspects of electromagnetic processes and beta
  decays involving halos. Nuclear processes are less transparent, but
  final state structures there should also be treated with great
  care. (Even though a resonance fit may describe the data it will not
  necessarily correspond to the physical reaction mechanism \cite{Gar01}.)
\item Can we find ways to experimentally study the hypernuclear halos ?
\item It would be very instructive to have a detailed experimental
  characterization of one of the halos in atomic and/or molecular physics.
\end{itemize}

The first decade or so of halo physics established the main concepts
and identified the main cases occuring among light nuclei. Apart from
a consolidation and refinement, reflected e.g.\ in the cross-checks on
the spatial extension that now have been successfully made in many
cases, the intervening years have also seen the concept being
established in molecular physics and the links to the Efimov effect
clarified. During the last years the first experiments at a new
generation of radioactive beam facilities have yielded new nuclear
cases and more experimental information should be available soon. This
may allow us to check our understanding of the conditions for halo
formation and thereby complete our picture of the phenomenon.

\ack
I am grateful to the many colleagues who during the past decades have
improved my understanding of halos and have delivered the impressive
experimental results reviewed here.
I would like to thank in particular Aksel Jensen and Bj\"{o}rn Jonson
for many enlightening discussions.

\end{document}